# Magneto-Optical Detection of Anisotropic Spin Currents in Altermagnetic $RuO_2$

Joongwon Lee[1,2,3], Jeonglyul Kim[4], Sreejith Nair[5], Seung Gyo Jeong[5], Changi Kim[1], Jae-Pil-So[6], Bohm-Jung Yang[1,7], Bharat Jalan[5], Hyobin Yoo[4,8], Farhan Rana[3]*, Taekoo Oh[6,9]*, and Hong-Gyu Park[1]*

[1]Department of Physics and Astronomy, and Institute of Applied Physics, Seoul National University, Seoul 08826, Republic of Korea.

[2]Research Institute of Basic Sciences, Seoul National University, Seoul 08826, Republic of Korea.

[3]School of Electrical and Computer Engineering, Cornell University, Ithaca, NY 14853, USA.

[4]Department of Materials Science and Engineering, Seoul National University, Seoul 08826, Republic of Korea.

[5]Department of Chemical Engineering and Materials Science, University of Minnesota, Twin Cities, Minneapolis, MN 55455, USA.

[6]Department of Physics, Soongsil University, Seoul 06978, Republic of Korea.

[7]Center for Theoretical Physics (CTP), Seoul National University, Seoul 08826, Republic of Korea.

[8]Research Institute of Advanced Materials, Seoul National University, Seoul 08826, Republic of Korea.

[9]Origin of Matter and Evolution of Galaxies (OMEG) institute, Soongsil University, Seoul 06978, Republic of Korea.

*Corresponding authors. Email: farhan.rana@cornell.edu; taekoo.oh@ssu.ac.kr; hgpark@snu.ac.kr

## Abstract

Altermagnets are a recently identified class of collinear antiferromagnets that host large spin-split electronic bands, offering a promising platform for efficient spin-current generation. Among proposed candidates, the metallic oxide $RuO_2$ is predicted to exhibit strong altermagnetic spin splitting; however, whether it sustains robust magnetic order beyond the ultrathin thickness limit remains unresolved. Here, we employ optical probes to investigate charge-to-spin conversion in a 12-nm-thick (101)-oriented $RuO_2$ film grown on sapphire. Polarization-resolved second-harmonic generation reveals nonlinear optical responses consistent with the surface symmetry and Néel order of $RuO_2$. Under an applied current, both second-harmonic generation and polar magneto-optical Kerr effect measurements detect a pronounced, directionally anisotropic spin polarization, exhibiting enhanced signals for current along [010] and strongly suppressed responses for current along $[\bar{1}01]$, in agreement with the symmetry of the altermagnetic spin-splitter effect. Non-magnetic or Rashba-type mechanisms cannot explain this symmetry-selective response. Scanning transmission electron microscopy further reveals that substantial strain persists even in relatively thick films, providing a possible explanation for the observed behavior. Therefore, these results establish $RuO_2$ as an efficient spin source and demonstrate the potential of altermagnets for field-free spintronic devices.

## INTRODUCTION

Altermagnets have recently emerged as a new class of antiferromagnets that exhibit large spin splitting on the order of eV despite their collinear magnetic order, attracting significant attention in condensed-matter physics [1-3]. In contrast to conventional collinear antiferromagnets, where opposite-spin bands remain degenerate, altermagnets host momentum-dependent spin splitting enforced by crystal symmetry, giving rise to spin-polarized Fermi surfaces without net magnetization. This distinctive electronic structure enables efficient spin-current generation mechanisms that are fundamentally different from conventional relativistic spin–orbit effects.

Among the proposed candidates, the metallic oxide $RuO_2$ is particularly attractive because of its high Néel temperature ($T_N$ ~ 500 K), large spin splitting, and high electrical conductivity, making it a promising platform for room-temperature spintronics [4-9]. $RuO_2$ crystallizes in the rutile structure, in which the Ru magnetic moments align along the [001] direction and are antiferromagnetically coupled (Fig. 1a) [4,5]. The local environment consists of anisotropic Ru–O octahedra rotated by 90° between neighboring sublattices, giving rise to symmetry-driven, momentum-dependent spin splitting. The resulting band structure underpins the altermagnetic spin-splitter effect (ASSE): an applied charge current shifts the Fermi contour, generating a spin current with polarization set by the Néel vector [10-12]. This contrasts with the spin Hall effect (SHE), in which spin polarization is orthogonal to the direction of charge and spin currents. In (101)-oriented $RuO_2$, the ASSE leads to a strongly anisotropic response, producing a predominantly out-of-plane spin current for current along [010] and an in-plane spin current for current along $[\bar{1}01]$.

Despite its potential to generate out-of-plane spin polarization, whether $RuO_2$ hosts robust long-range altermagnetic order beyond the ultrathin limit remains debated. Early studies using neutron diffraction and resonant X-ray scattering reported an antiferromagnetic order with a small moment (~0.05 $\mu_B$) at room temperature [4,5]. More recent investigations, however, have suggested that bulk $RuO_2$ − and possibly even films with thicknesses on the order of 10 nm − may exhibit much weaker, or even nearly negligible magnetic order, based on ARPES and muon-based measurements [13-16]. At the same time, electrical transport experiments on (101)-oriented $RuO_2$ films, including spin-torque ferromagnetic resonance and spin pumping studies, have reported symmetry characteristics broadly consistent with ASSE [6,8,11,12,17-20]. While these results provide evidence for current-induced spin conversion, disentangling altermagnetic responses from extrinsic contributions, including interfacial inhomogeneity or proximity effects, remains challenging.

Furthermore, epitaxial strain has emerged as a potentially decisive factor. In ultrathin epitaxial $RuO_2$ films (<4 nm) grown by hybrid molecular beam epitaxy (MBE), various experiments, including anomalous Hall effect (AHE) and optical second-harmonic generation (SHG), have suggested evidence of epitaxial strain stabilizing an altermagnetic order [9,15,16,21-24]. This raises an important question, as most prior reports of ASSE- or inverse ASSE-like behavior have been obtained in thicker films (> 6 nm), where strain is expected to relax. Another limitation is the lack of a direct, noninvasive probe capable of identifying ASSE and quantifying the intrinsic efficiency of spin-current generation without complications arising from interface transparency, spin backflow, or proximity effects. Establishing such a probe is essential for clarifying both the symmetry and the microscopic origin of spin generation in $RuO_2$.

Here, we address these challenges by employing two complementary optical probes, polarization-resolved SHG and high-sensitivity polar magneto-optical Kerr effect (MOKE), which directly access both magnetic symmetry and current-induced spin accumulation in a (101)-oriented $RuO_2$ film. Through these probes, we investigate the symmetry and magnitude of spin accumulation in a 12-nm-thick film. SHG reveals nonlinear optical responses associated with symmetry-breaking and in-plane components of surface spin accumulation, whereas polar MOKE, with a sensitivity of ~10 nrad and supported by electronic-structure calculations based on the Kubo formalism, provides a quantitative measure of the spin Hall angle and spin-current generation efficiency. Our measurements reveal a pronounced, symmetry-selective spin response consistent with ASSE and incompatible with conventional SHE. Furthermore, scanning transmission electron microscopy (STEM) analysis shows that substantial strain persists even in relatively thick films, depending on growth orientation and substrate, providing a clear explanation for the coexistence of strong spin-conversion signals and previously reported ASSE-like behavior in relatively thick $RuO_2$ films [6,8,11,12,17-19]. Together, these results establish $RuO_2$ as an efficient and symmetry-controlled spin source and highlight the broader potential of altermagnets for field-free spintronic applications.

## RESULTS

### Symmetry-governed anisotropic spin-current generation

We first consider the characteristic spin-split electronic structure underlying altermagnetic transport in $RuO_2$. As illustrated in Fig. 1b, the momentum-dependent spin splitting gives rise to spin-polarized Fermi contours with opposite spin textures related by crystal symmetry, leading to anisotropic spin-current generation. The allowed form of this spin response is further constrained by magnetic symmetry. In the presence of Néel order, the symmetry of the nonmagnetic $RuO_2$ (101) surface is reduced from the space group P4$_2$'/mnm' to the wallpaper group *pg*, leaving only the nonsymmorphic mirror $\{m_{010}| -1/2,1/2,1/2\}$ as nontrivial element (Fig. 1c). This symmetry imposes strict selection rules on spin generation: a current applied along $[\bar{1}01]$ preserves the mirror symmetry, whereas a current along [010] breaks it, allowing for a net out-of-plane spin accumulation.

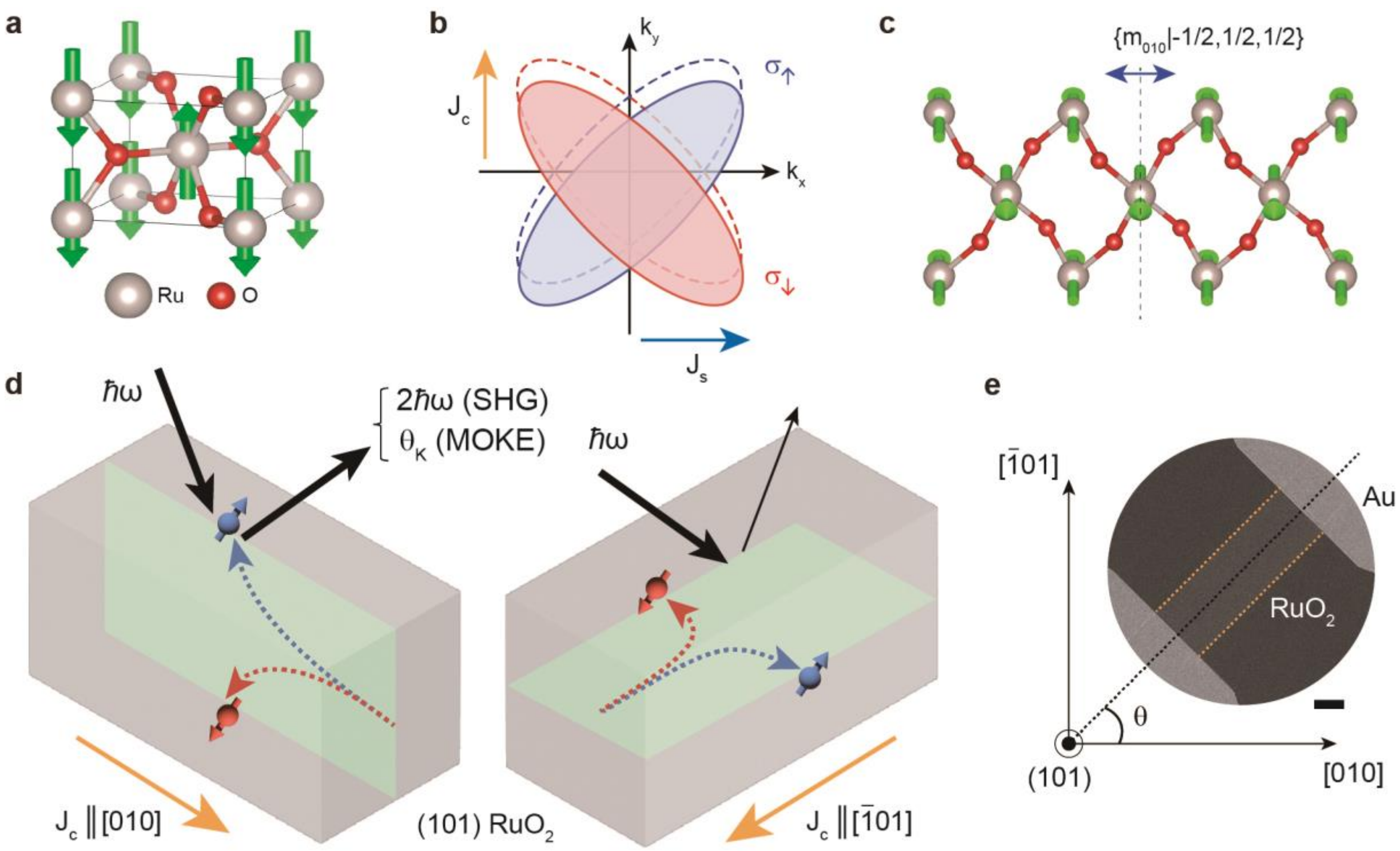

**Fig. 1. Crystal symmetry and optical detection scheme. a,** Crystallographic and magnetic structures of $RuO_2$. Gray and red spheres denote Ru and O atoms, respectively, and green arrows indicate the Ru magnetic moments. **b,** Schematic spin-split Fermi surfaces of $RuO_2$, showing spin-up ($\sigma_\uparrow$, blue) and spin-down ($\sigma_\downarrow$, red) states. The dotted and solid lines represent the equilibrium and current-shifted nonequilibrium distributions, respectively, for a current applied along [010]. **c,** Magnetic symmetry of the $RuO_2$ (101) surface, with the nonsymmorphic mirror plane $\{m_{010}|-1/2,1/2,1/2\}$ as the remaining symmetry operation. **d,** Optical probing geometry for charge-to-spin conversion in (101)-oriented $RuO_2$ under charge current applied along [010] (left) and $[\bar{1}01]$ (right). **e,** SEM image of a representative device, showing the definition of the angle θ with respect to the [010] direction. The $RuO_2$ film is enclosed by the orange lines. Scale bar, 10 μm.

Based on this symmetry consideration, we probed the current-direction dependence of spin accumulation using optical techniques (Fig. 1d). When a charge current flows along [010], the ASSE produces an out-of-plane spin accumulation that is optically detectable (Fig. 1d, left). In contrast, for the current applied along $[\bar{1}01]$, the generated spin current lies predominantly in-plane, and is therefore much less accessible in our optical measurements (Fig. 1d, right). To experimentally verify this anisotropy, we fabricated devices with different crystallographic orientations. Figure 1e shows a representative scanning electron microscopy (SEM) image, where the current direction is defined by an angle θ relative to [010], enabling systematic comparison of optical responses across different current directions.

The (101)-oriented $RuO_2$ film was grown on an *r*-plane $Al_2O_3$ substrate by molecular beam epitaxy (see Methods). The film thickness was chosen to exceed the spin diffusion length [25], thus suppressing optical contributions from the bottom interface and allowing high current densities. In addition, the *r*-plane $Al_2O_3$ substrate was selected because its high thermal conductivity [26] and wide band gap [27] help suppress Joule heating and minimize parasitic optical background signals. X-ray diffraction (XRD) and high-resolution transmission electron microscopy (HRTEM) confirmed the overall high crystallinity and the epitaxial relationship of the $RuO_2$ thin film (see Supplementary Note 1 and Fig. S1).

**SHG measurements of spin accumulation**

We investigate charge-to-spin conversion in $RuO_2$ using complementary optical probes: SHG and MOKE. In a normal-incidence reflection geometry, SHG is primarily sensitive to in-plane spin components, whereas MOKE selectively detects out-of-plane spin accumulation (see Supplementary Note 2). We first probed the in-plane spin accumulation induced by the ASSE using SHG. Figure 2a shows the measurement setup, which enables polarization-selective detection of nonlinear optical signals under applied DC current (see Methods). By controlling the polarizer (α) and analyzer (β) angles with respect to the [010] axis, we selectively accessed elements of the nonlinear susceptibility tensor $\chi^{(2)}$. The measured spatial map of the SHG signal confirms that the response is uniformly localized within the patterned $RuO_2$ region, with negligible background from the substrate (Fig. 2a, inset).

Polarization-resolved SHG measurements establish the structural and magnetic symmetry of the system [28-30]. As shown in Fig. 2b, the multi-lobed angular dependence of the SHG intensity reflects the symmetry of the (101)-oriented $RuO_2$ surface. The observed polarization patterns are consistent with magnetic-order-induced SHG (MSHG), whereas non-magnetic contributions, including bulk electric-magnetic-dipole SHG, fail to account for the data. The MSHG mechanism is described by

$$P_i(2\omega) = \left[\eta_{ijkl}L_l + \gamma_{ijkl}M_l\right]E_j(\omega)E_k(\omega).$$

Here, $L$ denotes the Néel vector and $M$ represents the magnetization. The indices $i$, $j$, $k$, $l$ denote the laboratory coordinates ($x$, $y$, $z$), aligned with the crystallographic directions [010], $[\bar{1}01]$, and [101], respectively (see Supplementary Note 3 and Fig. S2 for detailed symmetry arguments). Additional symmetry breaking from strain or surface reconstruction changes the MSHG interpretation only slightly.

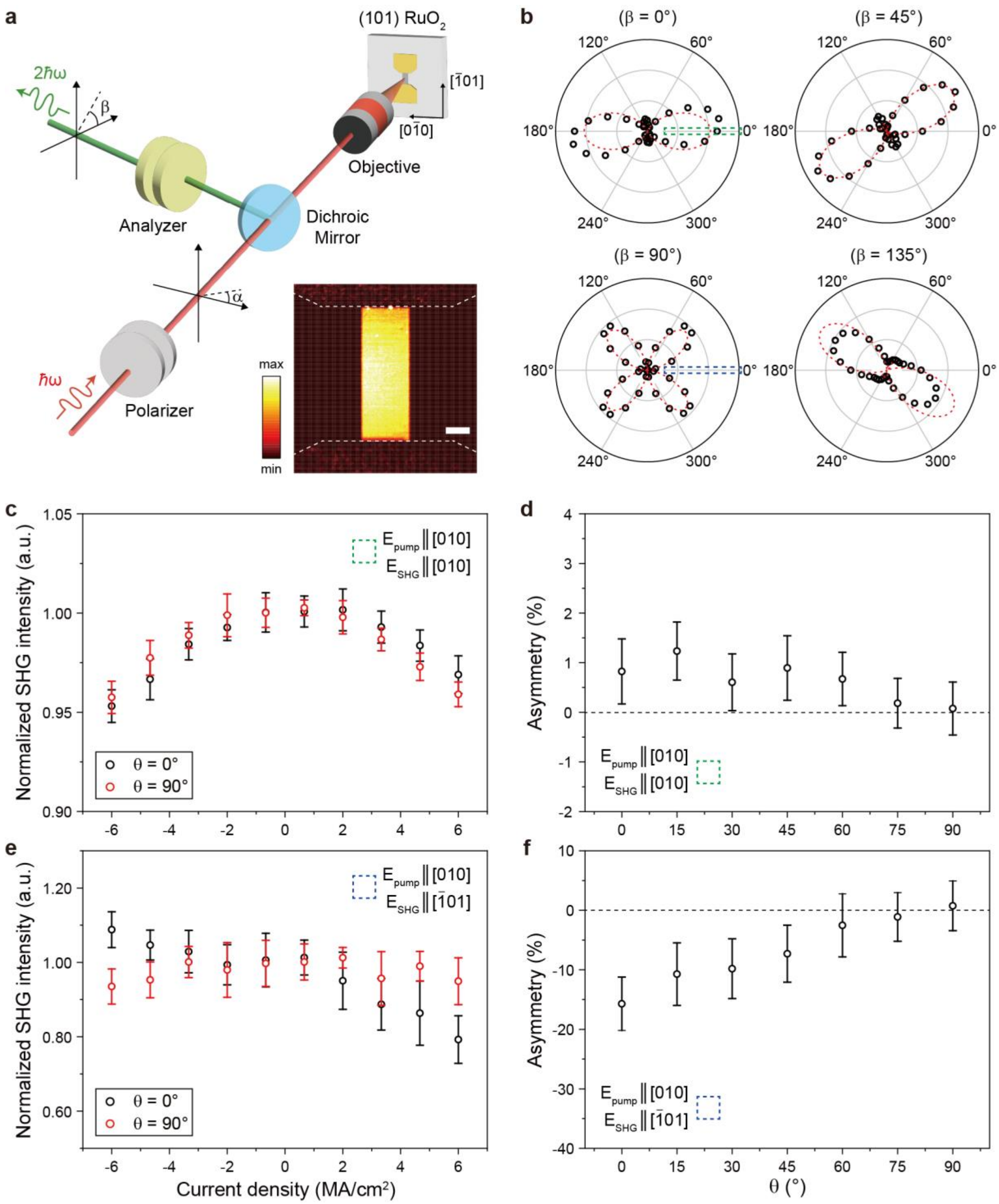

**Fig. 2. SHG measurements of Nèel order. a,** Schematic diagram of the SHG setup. A half-wave plate and a polarizer are used to control and analyze the polarization states of the incident fundamental beam and the emitted SHG signal, respectively. Inset: Scanning SHG image. Scale bar, 10 μm. **b,** Polarization-resolved SHG patterns for analyzer angles (β) of 0°, 45°, 90°, and 135°

with respect to the [010] axis of $RuO_2$. The azimuthal angle corresponds to the polarizer angle (α). Black circles represent the measured data, and the red dashed lines represent the fits. The SHG response at β = 0° exhibits a two-lobed polar pattern with maxima at α = 0°, whereas the pattern at β = 90° evolves into a four-lobed structure with maxima at α = ±45° and ±135°. Dashed boxes represent polarizer/analyzer configuration used in **c,d** (green) and **e,f** (blue). **c,e,** Normalized SHG intensity as a function of applied current for conduction channels along the [010] direction (θ = 0°) (**c**) and the $[\bar{1}01]$ direction (θ = 90°) (**e**), with both the polarizer and analyzer aligned along [010]. The data are normalized to the SHG intensity at zero applied current. **d,f,** SHG asymmetry as a function of θ. The polarizer is fixed along [010], and the analyzer is aligned along [010] (**d**) and $[\bar{1}01]$ (**f**).

We then examined the current dependence of the SHG intensity for currents applied along [010] (θ = 0°) and $[\bar{1}01]$ (θ = 90°) (Fig. 2c). This measurement reveals several key features. First, the SHG intensity decreases with increasing applied current. While this reduction is attributed to Joule heating, which suppresses the Néel order symmetrically with respect to current polarity, the observed asymmetry cannot be explained by thermal effects alone (see Supplementary Note 4 and Fig. S3). Instead, it can be understood within a surface-magnetization-induced SHG model [31], in which the nonlinear polarization generated by current-induced spin accumulation interferes constructively or destructively with the static SHG, depending on the current direction. Second, the SHG asymmetry appears only for current applied along [010] (θ = 0°) (Fig. 2c, black), whereas no measurable asymmetry is observed for current applied along $[\bar{1}01]$ (θ = 90°) (Fig. 2c, red). This directional dependence is consistent with the ASSE mechanism, in which the symmetry of the (101) surface governs the relationship between the applied charge current, the generated spin current, and the resulting spin accumulation.

To isolate the SHG contribution arising from spin accumulation from the Joule-heating-induced reduction of the Néel vector, we define the SHG asymmetry $A$ as

$$A = \frac{I(+J) - I(-J)}{I(+J) + I(-J)},$$

where $I(\pm J)$ denotes the SHG intensity measured under an applied current $J$ [31,32]. This definition exploits the different parity of the two effects: Joule heating is even in $J$, whereas ASSE-induced spin accumulation is odd. Accordingly, the numerator $I(+J) - I(-J)$ removes the symmetric thermal background and isolates the interference term between the static SHG component and the current-induced magnetization component. At $J = 6 \times 10^6$ A/cm$^2$, we plotted $A$ as a function of the current-direction angle θ, varied in 15° increments (Fig. 2d). $A$ decreases monotonically as θ increases from 0° to 90°, in agreement with the symmetry constraints of the ASSE.

Rotating the analyzer to $[\bar{1}01]$ provides access to orthogonal spin components along [010] (see Supplementary Note 2). Figure 2e shows the measured SHG intensity in this configuration. The static SHG signal is strongly suppressed, yet a finite current-induced signal is observed, possibly due to slight surface-induced canting of the Néel vector. When the current is applied along [010] (θ = 0°), the SHG intensity increases for negative current polarity despite the Joule-heating-induced suppression of SHG. In addition, the corresponding asymmetry is significantly enhanced (by a factor of ~15; Fig. 2f), indicating that even a small spin accumulation can produce a large relative response when the background is minimized. Importantly, the signal remains strongly anisotropic, appearing only for current along [010] and vanishing for current along $[\bar{1}01]$. This symmetry-selective response rules out a dominant contribution from the SHE and confirms that the observed spin accumulation originates from the ASSE [33].

**MOKE measurements of spin current**

SHG measurements provide a high-contrast probe of the symmetry of the current-induced spin accumulation but are less suitable for quantitative determination of the induced out-of-plane spin components. We therefore performed MOKE measurements at 4 K. Because the Kerr rotation induced by spin accumulation in metallic systems with short spin diffusion lengths is typically on the order of sub-microradians [34], we developed a high-sensitivity MOKE setup (Fig. 3a). The 4 K cryostat minimized temperature rise during current injection, enabling reliable quantitative analysis.

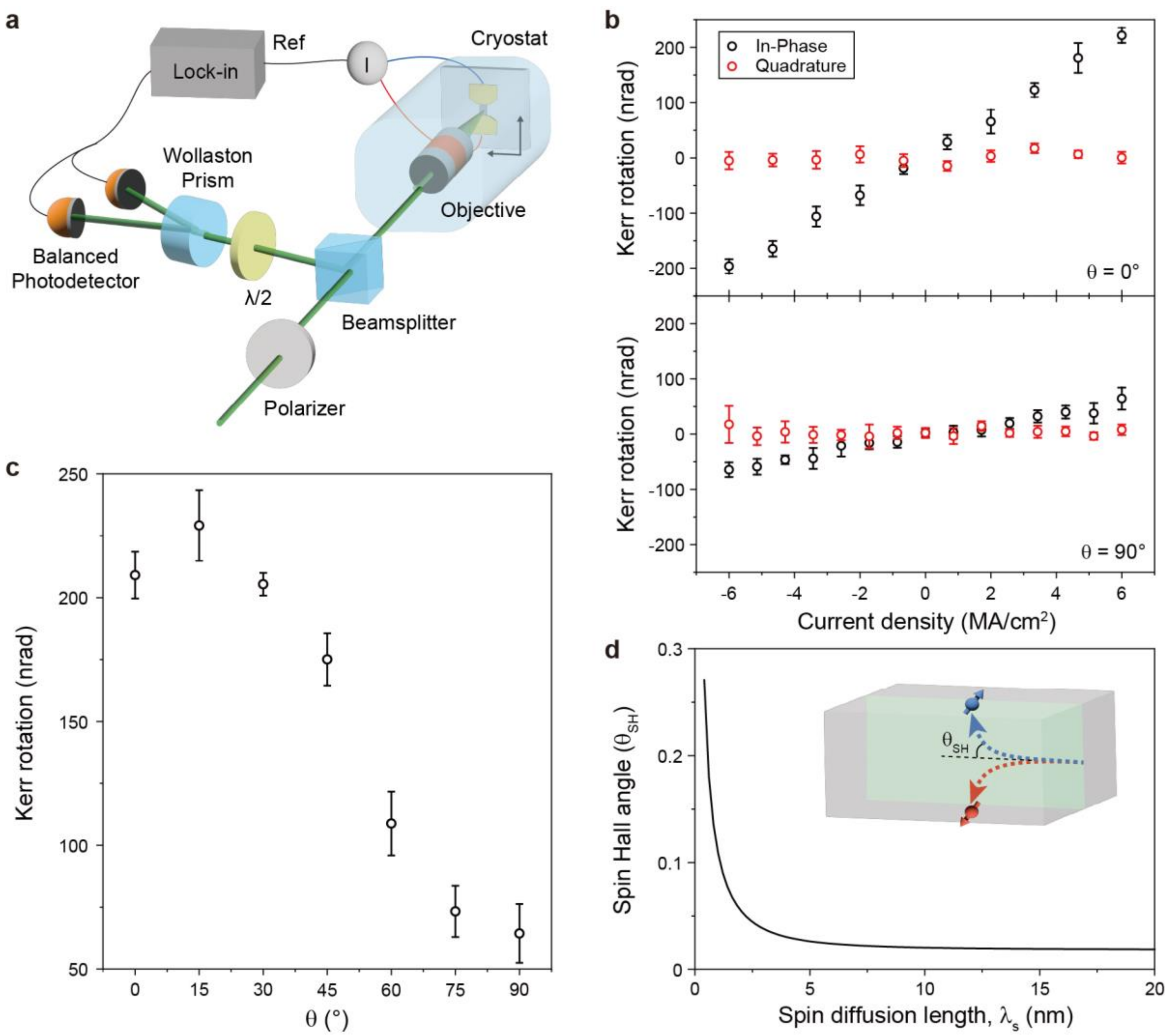

**Fig. 3. MOKE measurements of the ASSE. a,** Schematic of the MOKE setup. **b,** MOKE signal as a function of applied current for conduction channels along the [010] direction (θ = 0°, top) and the $[\bar{1}01]$ direction (θ = 90°, bottom). Black and red circles represent in-phase and quadrature

components, respectively. **c,** MOKE signal measured at a current density of 6 MA/cm$^2$ as a function of θ. **d,** Numerical analysis showing the relationship between the spin diffusion length ($\lambda_s$) and the spin Hall angle ($\theta_{SH}$) required to reproduce the measured spin accumulation.

In the setup, a continuous-wave (CW) laser at 532 nm was focused onto the active region of the device, while an AC current at $f$ = 2,337 Hz was applied. The reflected beam was analyzed using a half-wave plate, a Wollaston prism, a balanced photodetector, and a lock-in amplifier (see Methods). To calibrate the Kerr rotation, we determined the proportionality constant between the lock-in output and the Kerr rotation angle by rotating the half-wave plate. The wave plate was then fixed at the position that minimized the differential photocurrent before the measurements. To verify that the detected signal was phase-locked to the applied AC current, both the in-phase and quadrature components of the lock-in amplifier were monitored. In addition, by demodulating at the first harmonic ($f$) and tracking the phase, we distinguished genuine MOKE signals from Joule-heating-induced artifacts, which predominantly appear at the second harmonic ($2f$) (see Supplementary Note 5 and Fig. S4). By combining phase-sensitive detection with common-mode rejection, we achieved a MOKE sensitivity of ~27 nrad/$\sqrt{Hz}$, sufficient to resolve current-induced spin accumulation, as previously demonstrated in semiconductors [35,36] and heavy metals [34].

We then measured the MOKE signal while applying a charge current along different in-plane directions, characterized by the angle θ with respect to the [010] axis. Figure 3b shows the response to current applied along [010] (θ = 0°, top) and $[\bar{1}01]$ (θ = 90°, bottom). In both cases, the quadrature component (red dots) is negligible, whereas the in-phase component (black dots) varies linearly with the applied current, indicating that the signal is phase-locked to the excitation. Notably, the signal for current along [010] is approximately four times larger than that for $[\bar{1}01]$. To further investigate this anisotropy, we varied θ in 15° increments (Fig. 3c). The MOKE signal increases slightly, reaches a maximum near θ = 15°, and then decreases toward a minimum at θ = 90°. The full, current dependent in-phase and quadrature components for each angle are shown in Fig. S5. This trend indicates that both the magnitude and direction of the generated spin current are highly sensitive to device orientation.

These observations are consistent with the ASSE in $RuO_2$, which predicts strong anisotropy with respect to current direction. In metallic systems with strong spin–orbit coupling and short spin lifetimes, MOKE predominantly probes interfacial spin accumulation generated by out-of-plane spin currents [34]. Because the ASSE-induced spin current is perpendicular to both the applied charge current and the Néel vector, the observed angular dependence is consistent with an ASSE origin: the signal is enhanced for current along [010] and suppressed for current along $[\bar{1}01]$. In an idealized picture, projection of the current onto [010] would yield a cos θ dependence. The deviation from this behavior indicates that such simplified symmetry model fails to capture the response, likely reflecting the low symmetry of the $RuO_2$ (101) surface and finite incidence angles within the objective light cone, which can mix in-plane components into the nominally polar MOKE signal.

Next, we quantified the MOKE response based on the electronic structure of $RuO_2$. Using a 12-band model and the Kubo formula, we estimated the spin accumulation induced by the applied current [37]. In the low-field limit, the Kerr rotation per unit field is estimated to be ~0.122 rad/eV (see Supplementary Note 7 and Fig. S6). A measured Kerr rotation of 200 nrad corresponds to an effective magnetic field of ~141.8 G along [001], assuming a g-factor of 2 with a total spin J = 1/2. As the Oersted field (~2.57 G), corrected for the device geometry, accounts for only a negligible fraction of this effective field, we conclude that the spin accumulation is the dominant source of the observed Kerr rotation.

Using this calculated effective field, we evaluated the charge-to-spin conversion efficiency of the (101)-oriented $RuO_2$ films. The measured MOKE signal corresponds to a surface spin accumulation of $\delta\mu_0 \sim 1.64 \times 10^{-6}$ eV. To estimate the spin Hall angle ($\theta_{SH}$), we employed a 1D spin diffusion model [38]:

$$\theta_{SH} = \frac{\delta\mu_0}{J_c \rho \lambda_s \tanh(d/2\lambda_s)},$$

where $J_c$ is the current density, $\rho$ is the longitudinal resistivity, $\lambda_s$ is the spin diffusion length, and $d$ is the film thickness. Figure 3d shows the extracted $\theta_{SH}$ as a function of $\lambda_s$ for up to 20 nm. As expected, $\theta_{SH}$ and $\lambda_s$ exhibit a trade-off behavior. While $\theta_{SH} \sim 0.02$ for $\lambda_s \sim 12$ nm [25], it increases to ~0.1 for $\lambda_s \sim 1$ nm. Although these values are consistent with previous reports for rutile oxides [11,17], they remain nearly an order of magnitude smaller than the predicted value of ~0.317 [12]. This discrepancy may stem from multiple altermagnetic domains [39,40] (see Supplementary Note 7 and Fig. S6). Additionally, enhanced spin-flip scattering at the surface or due to epitaxial strain may reduce the effective spin diffusion length in (101)-oriented films, thereby suppressing the conversion efficiency.

**STEM imaging of strain**

Our optical studies of spin transport reveal two key features of charge-to-spin conversion in (101)-oriented $RuO_2$: (i) a symmetry in spin transport that is more consistent with the ASSE than the conventional SHE, and (ii) an experimentally extracted spin Hall angle that is an order of magnitude smaller than the theoretical prediction for the ASSE. To address this discrepancy, we investigated the nanoscale structural properties of the epitaxial films. Figure 4a presents a virtual annular dark-field (VADF) image reconstructed from 4D scanning transmission electron microscopy (4D STEM), with the *x*- and *z*- axes corresponding to the [010] and [101] crystallographic directions, respectively. The VADF image reveals non-uniform diffraction contrast within the $RuO_2$ layer, suggesting nanoscale structural heterogeneity arising from local variations in crystal orientation and/or phase, together with strain-induced lattice distortions.

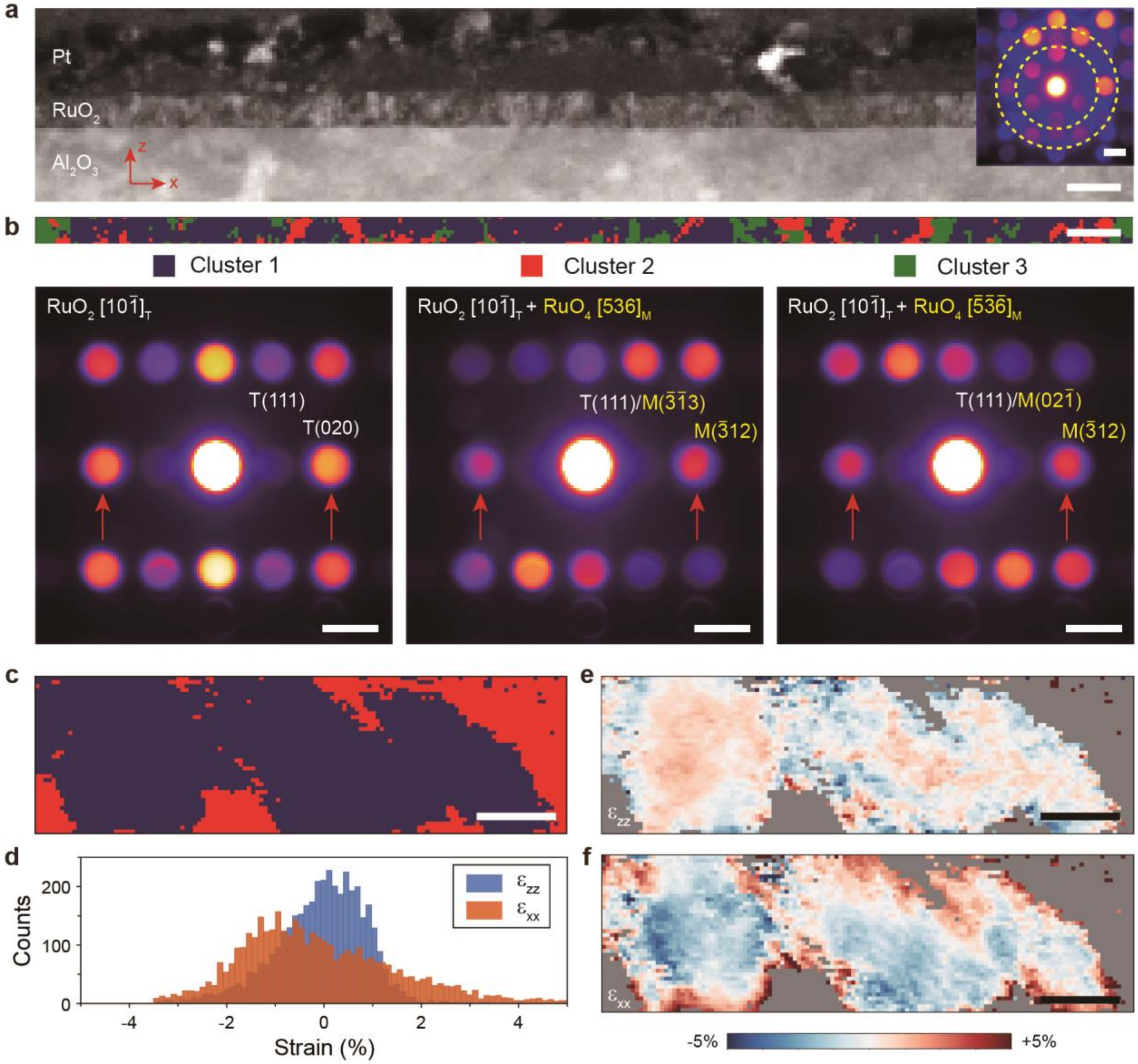

**Fig. 4. STEM imaging of thin-film strain. a,** Virtual annular dark-field (VADF) image generated from 4D STEM data. Scale bar, 20 nm. The inset shows an averaged NBED pattern, with the virtual annual detector indicated with a yellow annulus. Scale bar, 2 $nm^{-1}$. **b,** Spatial distribution map of the three structural clusters (Clusters 1–3; top) and their corresponding averaged NBED patterns (bottom). Scale bars, 20 nm (top) and 2 $nm^{-1}$ (bottom). **c,** Magnified cluster map obtained

from the $RuO_2$ thin film. Scale bar, 5 nm. **d,** Histograms of $\varepsilon_{zz}$ and $\varepsilon_{xx}$ measured from the Cluster 1 region shown in **c**. **e,f,** Strain maps of $\varepsilon_{zz}$ (**e**) and $\varepsilon_{xx}$ (**f**) obtained from the Cluster 1 region in **c**. Scale bar, 5 nm.

Figure 4b summarizes the structural analysis of the $RuO_2$ film. Based on the nanobeam electron diffraction (NBED) patterns acquired at each spatial pixel, we applied k-means clustering to classify the $RuO_2$ film into three primary structural clusters with distinct diffraction signatures. Notably, the diffraction disks indicated by the red arrows exhibit measurable shifts in Clusters 2 and 3 relative to the corresponding disks in Cluster 1. Similar shifts are also observed in neighboring diffraction disks, suggesting local structural variations within the $RuO_2$ film. Analysis of the NBED patterns indicates that Cluster 1 corresponds to the pristine (101)-oriented $RuO_2$ lattice, whereas Clusters 2 and 3 are tentatively assigned to distorted regions with distinct ruthenium oxide phases. The NBED patterns of Cluster 2 and 3 can be indexed using a monoclinic $RuO_4$ phase viewed along the [536] and [$\bar{5}\bar{3}\bar{6}$] zone axes, respectively (see Supplementary Note 8 and Fig. S7). The areal fractions of Clusters 1, 2, and 3 are approximately 69%, 15%, and 16%, respectively.

To further investigate the strain distribution in the $RuO_2$ thin film, we performed 4D STEM-based strain mapping. Figure 4c shows a high-spatial-resolution cluster map obtained from the $RuO_2$ film. Consistent with the diffraction pattern-based classification in Fig. 4b, this map spatially resolves the pristine (101)-oriented $RuO_2$ lattice region as Cluster 1 and the $RuO_x$-related monoclinic-like distorted region as Cluster 2.

Figure 4d presents histograms of the strain analysis performed within the region shown in Fig. 4c. The analysis was restricted to the pristine (101)-oriented $RuO_2$ lattice region, which constitutes the dominant phase in the film. Using the average diffraction pattern of this pristine region as the reference, the resulting strain maps represent local lattice distortions relative to the average pristine (101)-oriented $RuO_2$ lattice. While $\varepsilon_{zz}$ is centered near 0% with a relatively narrow distribution, $\varepsilon_{xx}$ exhibits a broader and more asymmetric distribution, indicating more pronounced spatial variation in the in-plane lattice distortion within the $RuO_2$ thin film. The spatial strain maps in Figs. 4e and 4f further reveal that larger $\varepsilon_{xx}$ values are mainly localized in the pristine (101)-oriented $RuO_2$ region adjacent to the distorted regions. This broadened $\varepsilon_{xx}$ distribution and the spatial localization of higher $\varepsilon_{xx}$ values indicate substantial nanoscale in-plane strain heterogeneity, likely associated with lattice accommodation near the boundaries between the pristine $RuO_2$ matrix and the distorted regions. Such nanoscale structural inhomogeneity provides a plausible structural origin for the observed spin-transport behavior, suggesting that strain-induced lattice distortions may play an important role in governing both spin generation and spin scattering.

**Conclusion**

In summary, our experiments provide signatures of spin transport in (101)-oriented $RuO_2$ using noninvasive optical probes, comprising SHG and high-sensitivity polar MOKE, with a focus on current-driven spin responses. Under applied current, both SHG and polar MOKE detect a spin polarization with strong directional anisotropy, exhibiting enhanced signals for current along [010] and strongly suppressed responses along $[\bar{1}01]$, consistent with the symmetry of the ASSE. In addition, polar MOKE measurements, with a sensitivity of ~10 nrad and supported by Kubo-formalism-based electronic-structure calculations, enable quantitative evaluation of the effective spin-torque efficiency. We further address the origin of the observed behavior in thick films by analyzing strain using STEM imaging.

We note that the observed anisotropic spin response cannot be readily explained by conventional mechanisms such as the SHE or the Rashba-Edelstein effect. Our current-induced SHG measurements indicate that SHE is not the dominant contribution, while the $C_{2v}$ symmetry in the nonmagnetic $RuO_2$ (101) surface effectively precludes Rashba-type spin splitting. Accordingly, the results are most consistent with the ASSE mechanism, in which the crystal's magnetic symmetry determines both the spin quantization axis and the direction of the generated spin current.

We also consider the implications of the structural heterogeneity identified via STEM. Recent reports suggest that epitaxial strain plays a key role in stabilizing the altermagnetic ground state in $RuO_2$ [9,21]. Although the 12-nm film thickness exceeds the expected strain-relaxation threshold, our 4D-STEM analysis reveals that significant strain persists within localized structural clusters. We propose two primary factors underlying the reduced spin Hall angle ($\theta_{SH}$) extracted from MOKE measurements: (i) the altermagnetic order, and hence the ASSE, may be confined to these strained nanodomains rather than uniformly distributed, and (ii) the high density of grain boundaries may act as spin-flip scattering centers, thereby shortening the effective spin diffusion length ($\lambda_s$) relative to previously reported values [25].

These results help clarify the microscopic origin of spin-current generation in $RuO_2$ and identify strain-stabilized structural regions as a possible condition for sustaining altermagnetic order beyond the ultrathin limit. Taken together, our findings establish $RuO_2$ as an efficient, symmetry-controlled spin source and highlight the potential of altermagnets for field-free spintronic devices.

## Methods

**Sample growth and characterization.** A 12-nm-thick (101) $RuO_2$ film was grown on an *r*-plane $Al_2O_3$ substrate by hybrid oxide molecular beam epitaxy (MBE) [41]. X-ray diffraction (XRD) scan and high-resolution transmission electron microscopy (HRTEM) image in Fig. S1 confirm the high crystallinity of the (101) $RuO_2$ film. The resistivity was measured to be 25 μΩ·cm at 4 K and 65 μΩ·cm at room temperature.

**Device fabrication.** The as-grown (101)-oriented $RuO_2$ film on *r*-plane $Al_2O_3$ substrate was capped in situ with an additional 2-nm $TiO_2$ layer. The device region ($20 \times 60$ $\mu m^2$) was patterned by photolithography followed by Ar ion milling. A second photolithography step was performed to remove the $TiO_2$ layer in selected regions and to form Ohmic contact to the contact pads. Subsequently, Ti (10 nm)/Au (100 nm) contact pads were deposited by liftoff.

**SHG measurements.** A Ti:sapphire laser (Coherent Chameleon Vision II) was used to excite the sample. The polarization states of the pump and second-harmonic (SH) light were controlled using a polarizer and an analyzer, each consisting of a half-wave plate and a polarizer. The SH light collected in a reflection geometry was directed to a monochromator/CCD system (Princeton Instruments PIXIS 400 BRX). The devices were driven using a Keithley 6221 current source.

**MOKE measurements.** MOKE measurements were performed using a 532 nm continuous-wave laser. The linear polarization of the excitation laser was defined using a polarizer. The laser beam was focused to a ~1 μm spot on the device area within a He-flow cryostat (Montana Instruments s50). The reflected light was directed through a beamsplitter into the detection path, which consisted of a half-wave plate, a Wollaston prism, and a balanced photodetector. The devices were driven using a Keithley 6221 AC current source, which served as the reference source for a Stanford Research SR830 lock-in amplifier. All MOKE measurements were performed at 4 K.

**Microstructural characterization.** The sample for cross-sectional transmission electron microscopy (TEM) was prepared using a focused ion beam (FIB) system (Helios 5 UC, Thermofisher Scientific). To minimize ion-beam-induced damage during the thinning process, Ga ions were sequentially applied at accelerating voltages of 30 kV and 5 kV, followed by a final thinning step performed at 2 kV. 4D scanning transmission electron microscopy (STEM) nanobeam electron diffraction (NBED) data were acquired using a Cs-corrected STEM (Themis Z, Thermofisher Scientific) operated at 300 kV with a probe semi-convergence angle of 1 mrad. NBED patterns were recorded at each probe position using an electron microscope pixel array detector (EMPAD) [42]. Local strain maps were generated using py4DSTEM by tracking the positions of Bragg disks across the scan and calculating their deviations from a reference lattice [43].

**Data availability.** The data that supports the findings of this study are available from the corresponding authors upon reasonable request.

**Code availability.** The codes used in this work are available from the corresponding authors upon request.

**Acknowledgements**

H.-G.P. acknowledges support from the Samsung Science and Technology Foundation (project no. SSTF-BA2401-02) and the National Research Foundation of Korea (NRF) grant funded by the Korean government (MSIT) (RS-2026-25468766). F.R. acknowledges support in part from SUPREME, one of seven centers in JUMP 2.0, a Semiconductor Research Corporation (SRC) program sponsored by DARPA. J.L. and F.R. acknowledge the use of the Cornell NanoScale Facility (CNF), a member of the National Nanotechnology Coordinated Infrastructure (NNCI), which is supported by the National Science Foundation (NSF Grant NNCI-2025233). T.O. was supported by the Pioneer Project for Future-Oriented Convergence Technology (Challenge Type; RS-2024-00416976) and Basic Science Research Program (RS-2025-16065011, RS-2026-25470523), through the National Research Foundation of Korea (NRF), funded by the Ministry of Science and ICT of Korean government, G-LAMP(RS-2025-25441317) through the National Research Foundation of Korea (NRF) funded by the Ministry of Education of Korean government. J.K. and H.Y. were supported by the National Research Foundation of Korea (NRF) grant funded by the Korea government (Ministry of Science and ICT, MSIT) (Grant No. RS-2021-NR060087, RS-2026-25491476). Film synthesis and structural characterizations (S.N, S.G.J. and B.J.) were supported by the U.S. Department of Energy through grant Nos. DE-SC0020211. S.G.J, and B.J. also benefited from the Air Force Office of Scientific Research Multi University Research Initiative (AFOSR MURI, Award No. FA9550-25-1-0262). Parts of this work were carried out at the Characterization Facility, University of Minnesota, which receives partial support from the NSF through the MRSEC program under Award No. DMR-2011401. S.N. acknowledges partial support from the NSF through the MRSEC program under Award No. DMR-2011401. Film growth was performed using instrumentation funded by AFOSR DURIP awards FA9550-18-1-0294 and FA9550-23-1-0085.

**Author contributions**

J.L., F.R., and H.-G.P. conceived the idea and directed the project. S.N., S.G.J., and B.J. synthesized the $RuO_2$ thin films via MBE and performed the initial structural characterizations using X-ray diffraction. J.L. fabricated the spintronic devices. J.L., C.K., and J.S. performed the magneto-optical measurements. T.O. and B.-J.Y. developed the theoretical models and performed the computational MOKE analysis. J.K. and H.Y. conducted the STEM-based structural and strain analyses. J.L. drafted the manuscript, with major contributions from H.-G.P. All authors discussed the results, contributed to the revision of the text, and approved the final version of the manuscript.

**Competing interests**

The authors declare no competing interests.

**Correspondence and requests for materials** should be addressed to F.R., T.O., and H.-G.P.

## Supplementary Note

### 1. Structural characterization

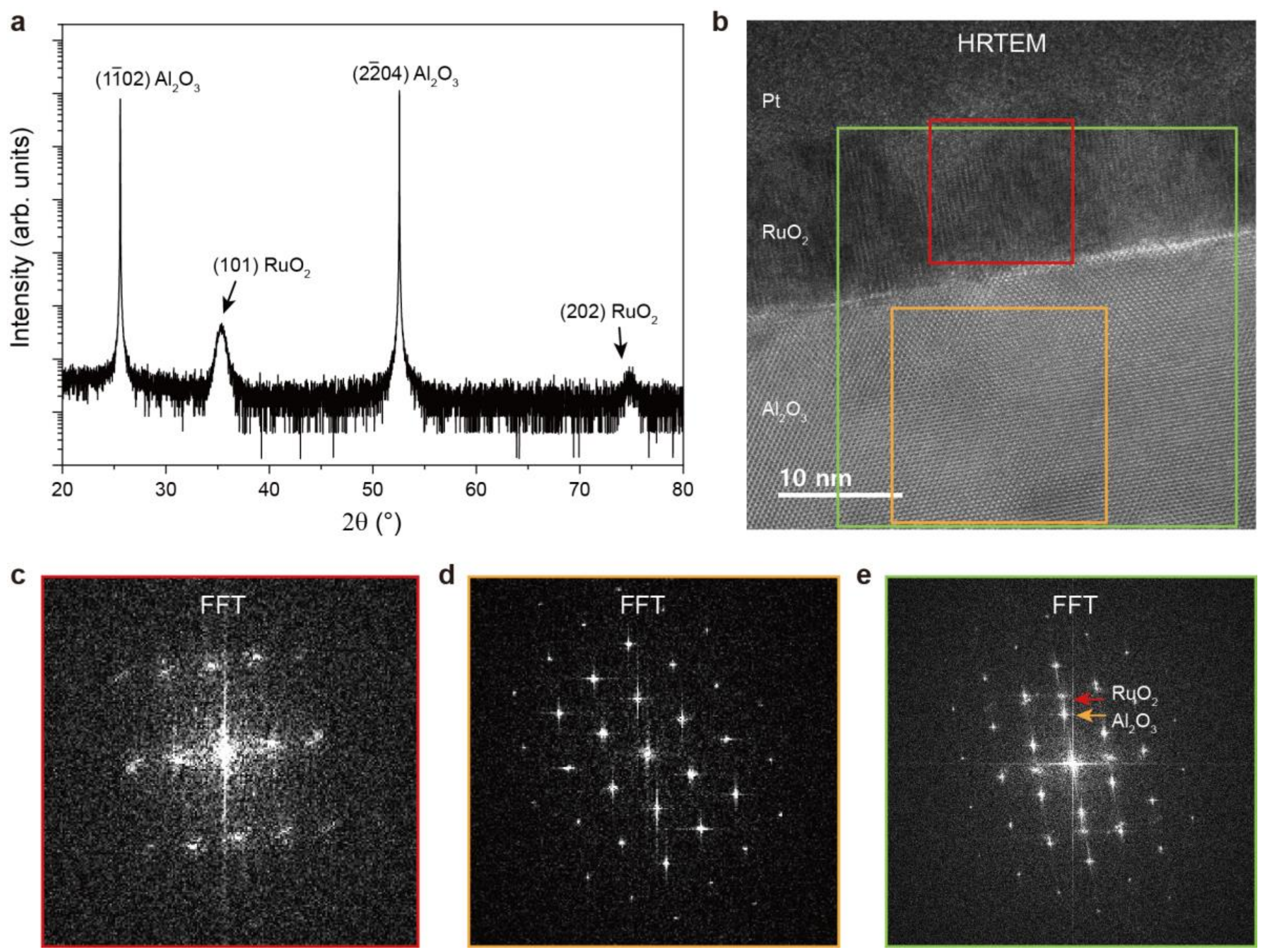

**Fig. S1. Structural characterization of the $RuO_2$ film. a,** θ–2θ X-ray diffraction (XRD) scan of a $RuO_2$ film grown on an *r*-plane sapphire substrate. **b,** Cross-sectional high-resolution transmission electron microscopy (HRTEM) image of the heterostructure. **c-e,** Fast Fourier transform (FFT) patterns corresponding to the color-coded regions in **b**.

The phase purity and epitaxial alignment of the $RuO_2$ films were first confirmed via X-ray diffraction (XRD). The θ–2θ scan displays a prominent reflection at 2θ ~ 35°, corresponding to the rutile $RuO_2$ (101) plane, alongside the *r*-plane sapphire $(1\bar{1}02)$ substrate peak at 2θ ~ 26° (Fig. S1a). Higher-order reflections, specifically the $Al_2O_3$ $(2\bar{2}04)$ at ~53° and the $RuO_2$ $(2\bar{2}04)$ at

~74°, further substantiate the phase-pure nature of the growth. The absence of additional rutile orientations or impurity phases confirms that the (101) growth direction is maintained throughout the film thickness, providing a robust structural foundation for the subsequent altermagnetic transport studies.

Atomic-scale insights into the heterostructure were obtained through cross-sectional high-resolution transmission electron microscopy (HRTEM). The micrograph in Fig. S1b clearly resolves the distinct layers of the Pt capping, the $RuO_2$ film, and the $Al_2O_3$ substrate, revealing a sharp interface. To evaluate the epitaxial relationship, fast Fourier transform (FFT) analysis was performed on the regions color-coded in Fig. S1b. The FFT of the substrate region (Fig. S1c) and that of the $RuO_2$ region (Fig. S1d) both exhibit sharp, well-defined diffraction spots, confirming high crystallinity. Furthermore, the FFT of the interfacial region (Fig. S1e) shows a clear superposition of both lattices with matching reciprocal-space vectors, directly demonstrating the epitaxial alignment between the (101)-oriented $RuO_2$ film and the *r*-plane sapphire substrate.

**2. SHG and MOKE tensors for surface-normal incidence**

In isotropic materials possessing $C_{2v}$ symmetry, the linear and nonlinear magneto-optical responses are governed by

$$\vec{P}(\omega) = \varepsilon_0 \begin{pmatrix} \varepsilon_{xx} & iQM_z & -iQM_y \\ -iQM_z & \varepsilon_{yy} & -iQM_x \\ iQM_y & -iQM_x & \varepsilon_{zz} \end{pmatrix} \vec{E},$$

$$\vec{P}(2\omega) = \varepsilon_0 \begin{pmatrix} \eta_{xxxy}M_y & \eta_{xyyy}M_y & \eta_{xzzy}M_y & \eta_{xyzz}M_z & \chi_{xxz} & \eta_{xxyx}M_x \\ \eta_{yxxx}M_x & \eta_{yyyx}M_x & \eta_{yzzx}M_x & \chi_{yyz} & \eta_{yxzz}M_z & \eta_{yxyy}M_y \\ \chi_{zxx} & \chi_{zyy} & \chi_{zzz} & \eta_{zyzx}M_x & \eta_{zyzy}M_y & \eta_{zxyz}M_z \end{pmatrix} \vec{E}\vec{E}.$$

Under a surface-normal geometry for both SHG and MOKE characterization of surface spin accumulation, SHG is primarily sensitive to in-plane spin components ($M_x$, $M_y$), whereas MOKE probes the out-of-plane component ($M_z$). However, the altermagnetic phase of the (101)-oriented $RuO_2$ surface necessitates a reduction in symmetry from $C_{2v}$ to $C_s$, with the mirror plane along the (010) remaining as the sole symmetry operation. Under this reduced symmetry, the SHG and MOKE tensors allow mixing between in-plane and out-of-plane components. Furthermore, epitaxial strain-induced symmetry breaking may introduce additional tensor elements, further complicating the analysis. In the present work, these higher-order cross-coupling contributions are neglected, as their magneto-optical coupling strengths are expected to be small relative to the primary terms and are difficult to quantify experimentally.

### 3. Static SHG characterization and symmetry analysis

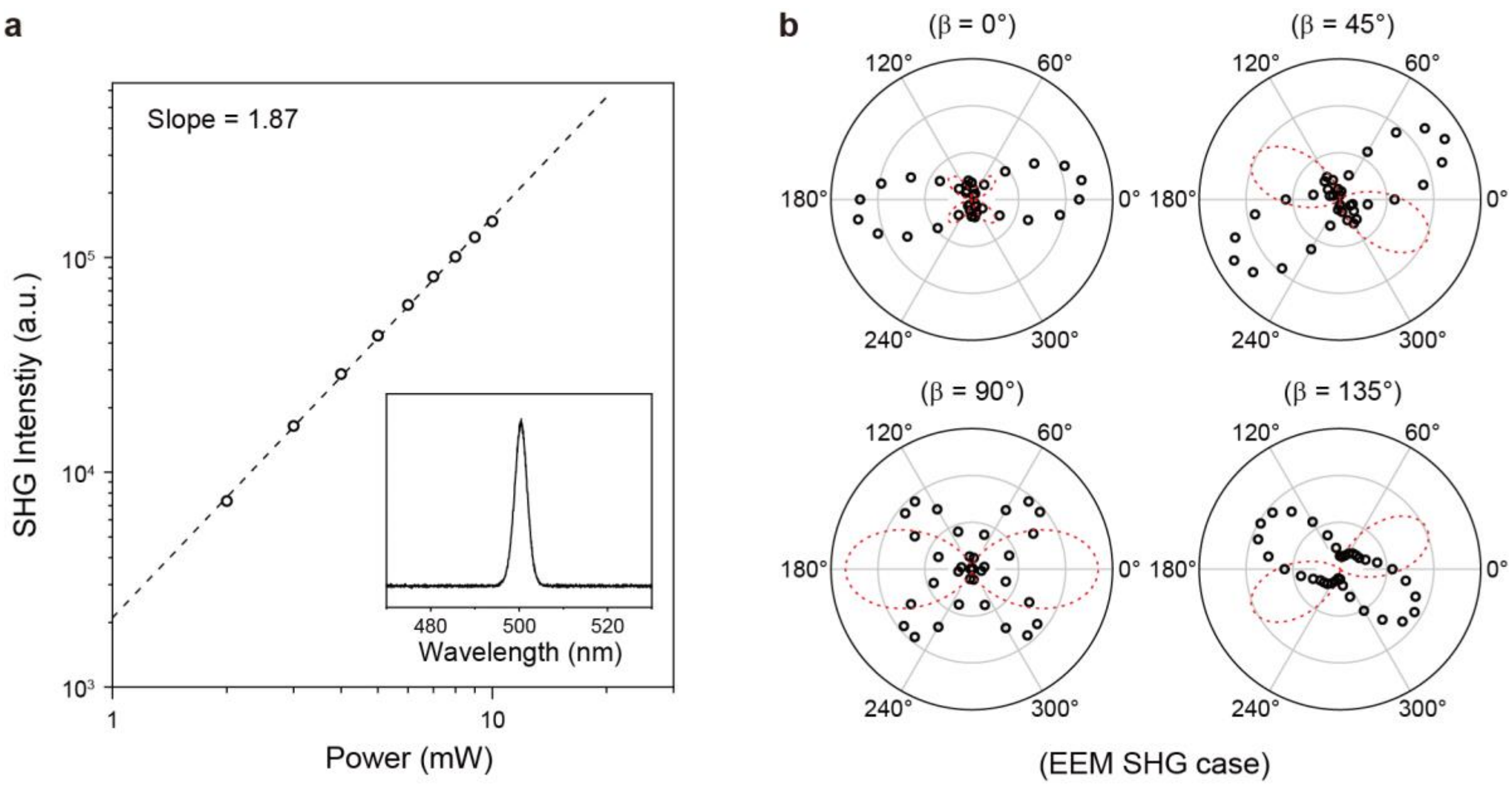

**Fig. S2. Static SHG response. a,** Power-dependent SHG intensity. Inset: SHG spectrum. **b,** SHG polar patterns for analyzer angles of 0°, 45°, 90°, and 135° with respect to the [010] axis of $RuO_2$. The azimuthal angle corresponds to the polarizer angle. Black circles represent the measured data, and the red dashed line represents a fit based on bulk electric-magnetic dipole induced SHG (EEM-SHG).

In this section, we provide additional details on the origin of SHG and the derivation of symmetry-allowed SHG components. Figure S2a shows the SHG intensity as a function of incident fundamental laser power, plotted on a log-log scale. A linear fit yields a slope of 1.87, consistent with the expected quadratic power dependence of a second-order nonlinear optical process. The inset shows a representative SHG spectrum centered at 500 nm, confirming that the observed signal arises from SHG, with negligible background contributions from photoluminescence or higher-order processes.

Next, we derive the symmetry-allowed SHG components for the MSHG case. In the altermagnetic phase of the (101)-oriented $RuO_2$ surface, the $C_s$ point group symmetry with a mirror reflection along (010), constrains nonzero components of the tensor $\eta_{ijkl}L_l$ to take the form:

$$P_i = \eta_{ijkl}E_jE_kL_l = \begin{pmatrix} \eta_{xxxy}L_y + \eta_{xxxz}L_z & \eta_{xyyy}L_y + \eta_{xyyz}L_z & \eta_{xxyx}L_x \\ \eta_{yxxx}L_x & \eta_{yyyx}L_x & \eta_{yyxy}L_y + \eta_{xxxz}L_z \end{pmatrix} \begin{pmatrix} E_x^2 \\ E_y^2 \\ 2E_xE_y \end{pmatrix}.$$

The tensor coupling to the magnetization, $\gamma_{ijkl}M_l$, follows the same symmetry form. Under these constraints, and assuming $\vec{L} \parallel [001]$, terms containing $L_x$ are negligible. The experimental data presented in Fig. 2b can be reproduced using the following relations:

$$\eta_{xyyy}L_y + \eta_{xyyz}L_z = -2.07 \times (\eta_{xxxy}L_y + \eta_{xxxz}L_z),$$

$$\eta_{yyxy}L_y + \eta_{xxxz}L_z = -0.46 \times (\eta_{xxxy}L_y + \eta_{xxxz}L_z).$$

Last, we explain why the EEM-SHG mechanism, described by $P_i(2\omega) = \chi_{ijk}^{eem}E_j(\omega)H_k(\omega)$, fails to account for our experimental observations. Under the $P4_2'/mnm'$ magnetic space group, to which bulk altermagnetic $RuO_2$ belongs, the nonzero tensor components are

$$\chi_{yzx}^{eem} = \chi_{zxy}^{eem} = \chi_{xzy}^{eem} = \chi_{zxy}^{eem},$$

$$\chi_{xyz}^{eem} = \chi_{yxz}^{eem},$$

where $x$, $y$, $z$ denote the crystal $a$, $b$, $c$ axes, respectively. To evaluate their contribution, we perform a coordinate transformation:

$$\chi_{ijk}^{eem,film} = \det(R)\, R_{ii'} R_{jj'} R_{kk'} \chi_{i'j'k'}^{eem},$$

where $R$ is the transformation matrix mapping the crystal axes to the film coordinates ([010], [$\bar{1}01$], [101]). The nonzero tensor components transform as follows (showing only components without the $z$ index):

$$\chi_{xyz}^{eem} \rightarrow \chi_{yxy}^{eem,film},$$

$$\chi_{yxz}^{eem} \rightarrow \chi_{xyy}^{eem,film},$$

$$\chi_{xzy}^{eem} \rightarrow \chi_{yyx}^{eem,film},$$

$$\chi_{zxy}^{eem} \rightarrow \chi_{yyx}^{eem,film},$$

$$\chi_{yzx}^{eem} \rightarrow \chi_{xyy}^{eem,film},$$

$$\chi_{zyx}^{eem} \rightarrow \chi_{yxy}^{eem,film}.$$

Symmetry analysis of the transformed tensor indicates that $y$-polarized SHG should be dominant under both $x$- and $y$-polarized excitation. In contrast, $x$-polarized SHG should vanish under purely $x$- or $y$-polarized pumping, requiring a simultaneous presence of both components. This predicted symmetry does not reproduce our experimental observations, as illustrated by the representative fit in Fig. S2b. Furthermore, given that the coupling strength of EEM-SHG is typically orders of magnitude weaker than that of electric-dipole SHG, we conclude that the measured SHG signal originates from MSHG and directly probes the surface Néel order.

### 4. Current-polarity dependence of SHG response

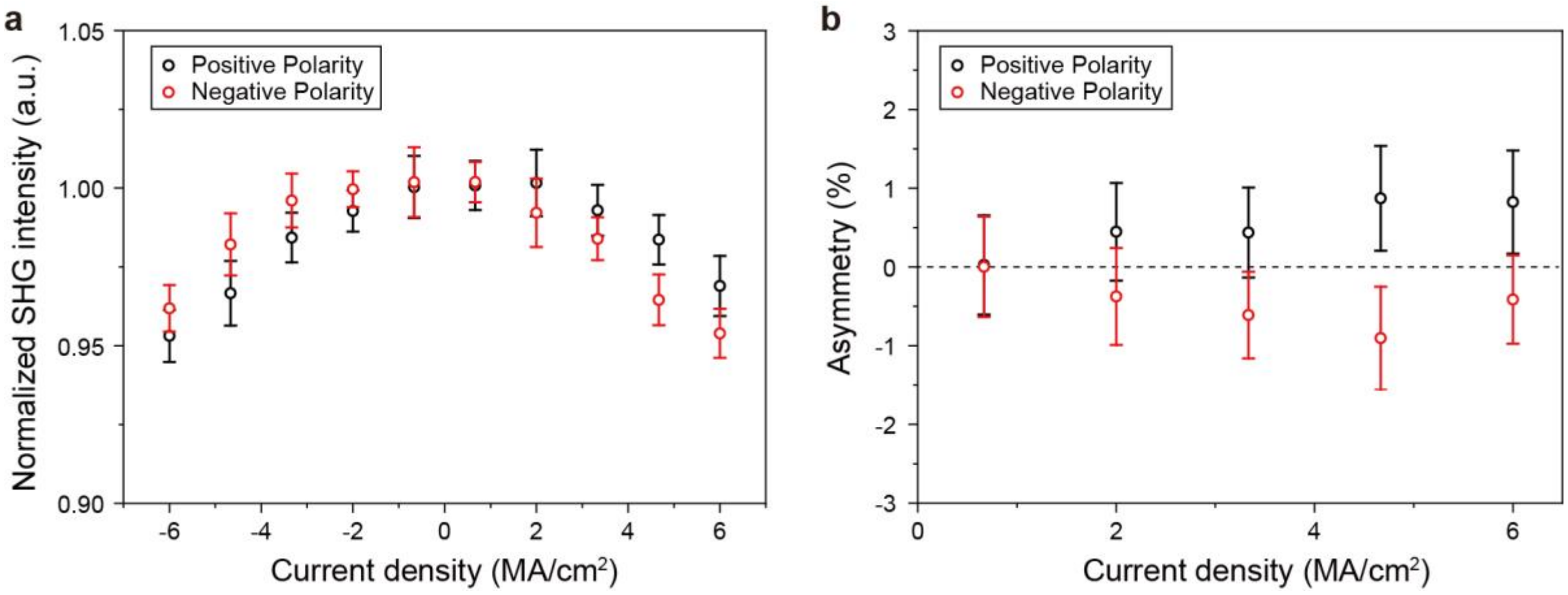

**Fig. S3. Polarity-dependent SHG response. a,b,** Normalized SHG intensity (**a**) and SHG asymmetry $A$ (**b**) measured under two opposite current polarities (positive and negative). Black and red circles represent the positive and negative polarities, respectively.

In this section, we demonstrate that the observed asymmetric change in SHG intensity under an applied current is intrinsic to the material response and does not originate from thermal gradients or circuit-induced offsets. To isolate the current-induced SHG signal from potential Joule heating or thermo-optical effects, we measured the normalized SHG intensity under two electrical configurations: positive and negative current polarity. In the negative polarity configuration, the source and ground leads were physically swapped at the device terminals.

As shown in Fig. S3a, the SHG intensity under negative polarity exhibits a mirrored dependence relative to that under positive polarity, confirming that the signal follows the direction of the charge current rather than a scalar thermal effect. Figure S3b presents the SHG asymmetry, $A$, as a function of the applied current. Although the linear scaling of $A$ with current is partially limited by experimental uncertainty at low signal levels, the clear sign reversal of the asymmetry upon reversing the current polarity confirms the nonreciprocal nature of the observed magneto-optic response.

## 5. Lock-in measurements to isolate intrinsic MOKE signals from Joule heating effects

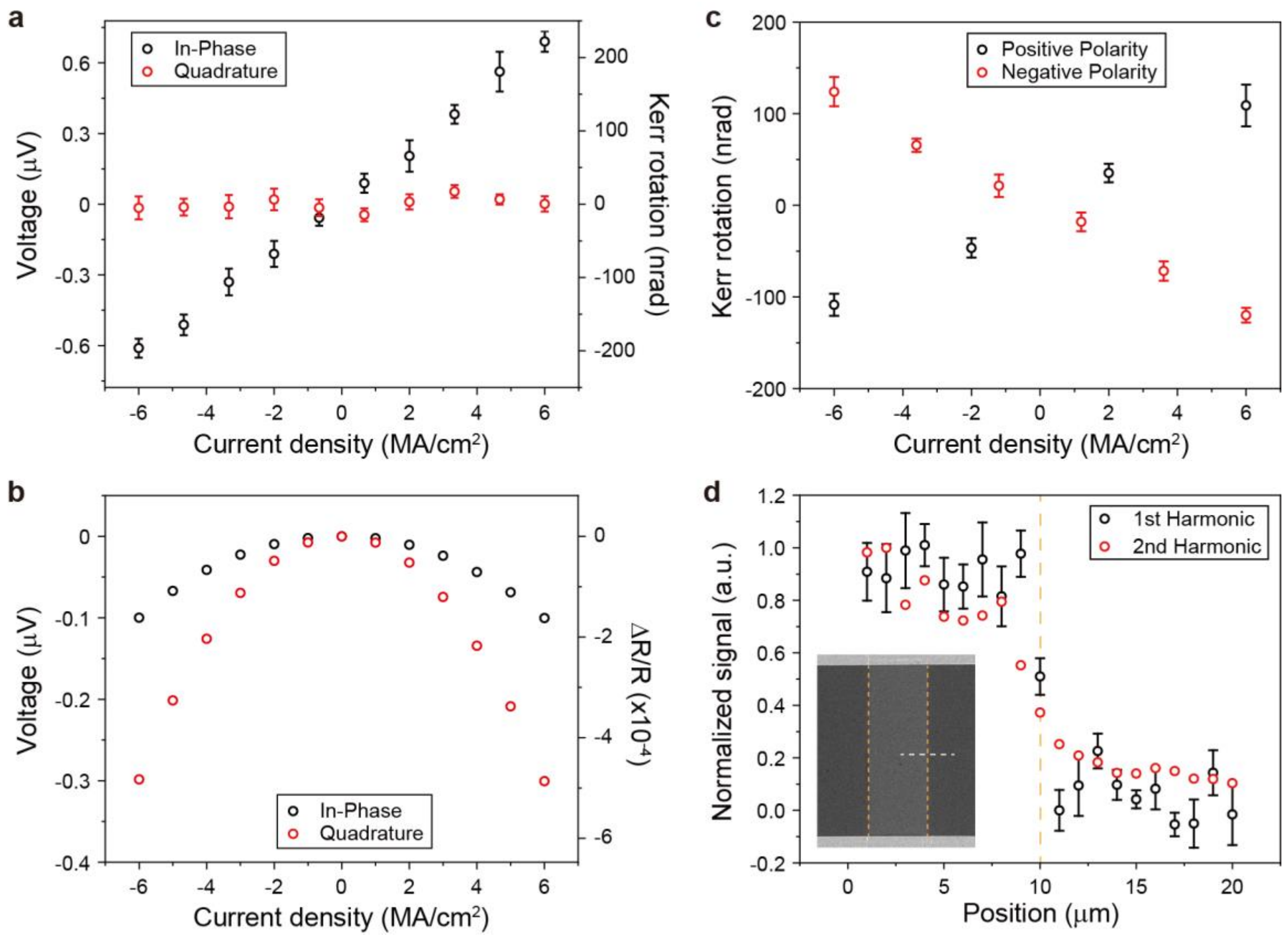

**Fig. S4. Lock-in detection of MOKE signals and separation of thermal contributions. a,b,** Measured first-harmonic ($f$) (**a**) and second-harmonic ($2f$) (**b**) lock-in voltage signals as a function of the applied charge current density ($J_c$). The primary $y$-axes show the raw lock-in output, while the secondary $y$-axes on the right provide the corresponding Kerr rotation (**a**) and relative reflectance change (**b**). Black and red circles represent the in-phase and quadrature components, respectively. **c,** Current-induced Kerr rotation measured under two opposite source configurations (positive and negative polarity). Black and red circles represent positive and negative polarities, respectively. **d,** Spatial line scan across the device boundary. Black and red circles represent the $f$ and $2f$ signals, respectively.

In this section, we address thermal artifacts that mimic current-induced MOKE signals – a challenge well documented since the earliest optical studies of the spin Hall effect in Pt and W [1-3]. In those systems, Joule-heating-induced reflectance changes often manifested as spurious

MOKE signals. While AC lock-in detection is commonly employed to mitigate such effects [4,5], a more rigorous decoupling is required to ensure the intrinsic nature of the altermagnetic signal.

To rule out thermal artifacts, we examined the raw lock-in voltage signals as a function of current density by demodulating the signal at the first (*f*) and second (2*f*) harmonic signals. Figures S4a and S4b show the *f* and *2f* signals measured directly from the lock-in output. For clarity, the dual *y*-axis on the right-hand side of the plots provides the corresponding Kerr rotation and relative reflectance change ($\Delta R/R$), respectively. The *f* signal exhibits a strictly linear dependence on the applied current, reaching an amplitude of approximately ±0.6 μV at ±6 × $10^6$ A/cm$^2$. In contrast, the 2*f* signal, which tracks the thermoreflectance, shows a quadratic dependence on current and is significantly larger, reaching ~30 μV at the same current level.

The common-mode signal of our balanced photodetector, $V_{PD}(t)$, is given by the product of the modulated intensity and the effective polarization angle:

$$V_{PD}(t) \propto P_0 R\big(1 + \Delta R(I)\big)\big(\phi_0 + \phi_1(I)\big),$$

where $P_0$ is the optical power, $R$ is the static reflectance, $\Delta R$ is the current-induced reflectance change, $\phi_0$ is the setup-dependent angular mismatch, and $\phi_1$ is the current-induced Kerr rotation. We model the thermoreflectance ($\Delta R$) as quadratic in current and the Kerr rotation ($\phi_1$) as linear, using proportionality constants $C_R$ and $C_\phi$:

$$\Delta R = C_R I^2, \phi_1 = C_\phi I.$$

For an applied sinusoidal current, $I(t) = I_0 \sin 2\pi f t,$ the expansion of $V_{PD}(t)$ yields signals at *f* and 2*f*:

$$V_f \propto P_0 R C_\phi I_0 \sin 2\pi f t,$$

$$V_{2f} \propto -\frac{1}{2} P_0 R C_R \phi_0 I_0^2 \cos 4\pi f t.$$

This linear dependence in the $f$ signal and the quadratic dependence in the $2f$ signal are consistent with the experimental observation in Figs. S4a and S4b.

While the $f$ signal is ideally purely magnetic, a small DC current offset ($I_1$), often arising from the device asymmetry [2], can create a “ghost” thermal contribution at the first harmonic. Assuming $I(t) = I_1 + I_0 \sin 2\pi ft$, the $f$ signal becomes:

$$V_{1\omega} \propto P_0 R\left(C_\phi I_0 + 2C_R \phi_0 I_1 I_0\right) \sin 2\pi f t.$$

The first term corresponds to the true MOKE signal, and the second represents the ghost thermal signal. To evaluate the risk of this ghost signal manifesting as MOKE signal, we compared the measured signal with the current asymmetry in the source and circuit. For an applied current of 14.4 mA, we observed a worst-case DC offset of $I_1 \sim 0.1$ μA. Using the above expression, the ratio of the thermal “ghost” leakage to the measured $2f$ signal is $4(I_1/I_0) \sim 2.8 \times 10^{-5}$. This limits the maximum thermal artifact in the $f$ signal to ~8 nV – nearly two orders of magnitude smaller than our observed 0.6 μV signal.

The origin of these signals is further clarified by their phase relative to the driving AC current. The $f$ signal remains purely in-phase, whereas the $2f$ signal is dominated by the quadrature component, which is approximately three times larger than the in-phase component. This 72° phase lag in the $2f$ channel is a characteristic of thermal diffusion, where the modulation frequency exceeds the thermal relaxation rate of the $RuO_2$ film and substrate interface. If the $f$ signal were of thermal origin, it would exhibit a similar phase lag. The absence of such a quadrature component at $f$ definitively rules out thermal contributions and confirms that the measured MOKE signal reflects instantaneous current-induced spin accumulation.

Figure S4c presents the current-induced Kerr rotation measured under two source configurations with opposite polarity. A complete and symmetric reversal of the signal is observed upon swapping the source and ground leads. This parity under current reversal is a fundamental requirement for the ASSE and confirms that the signal is coupled to the direction of the charge current.

To further distinguish the localized electronic response from the extended thermal background, we performed a spatial line scan across the device boundary (Fig. S4d). The $f$ and $2f$ signals exhibit markedly different spatial profiles. The first-harmonic signal remains uniform within the patterned $RuO_2$ region and drops sharply at the device edge, consistent with a signal tied to local current density and interfacial spin accumulation. In contrast, the $2f$ signal decays gradually and persists up to ~10 μm into the non-conductive substrate. Fluctuations in the $2f$ signal within the device region are attributed to local variations in reflectance arising from surface roughness or microscopic defects.

The persistence of the $2f$ signal beyond the device edge is consistent with the high thermal diffusivity of sapphire. Unlike spin accumulation, which is confined to the current path (within the spin diffusion length), Joule heat acts as a line source that diffuses laterally into the substrate. At the modulation frequency used here, the thermal diffusion length is calculated to be $\Lambda_{th}$ ~ 31 μm [6], providing a clear mechanism for the extended spatial tail in the $2f$ scan but absent in the $f$ scan.

**6. Raw MOKE signals from devices with various orientations**

The full current-dependent in-phase and quadrature components for each angle are shown in Fig. S5. While ASSE predicts no spin accumulation for current applied along the $[\bar{1}01]$ direction, we observe a suppressed yet nonzero signal for the $[\bar{1}01]$ direction (θ = 90°). This may partially stem from the rutile tetragonal structure of $RuO_2$, where the $[\bar{1}01]$ direction is not strictly orthogonal to the principal crystallographic axes, potentially allowing a residual ASSE contribution. In addition, the reduced symmetry of the (101) surface and finite numerical aperture effects can lead to leakage of in-plane spin components into the polar MOKE signal.

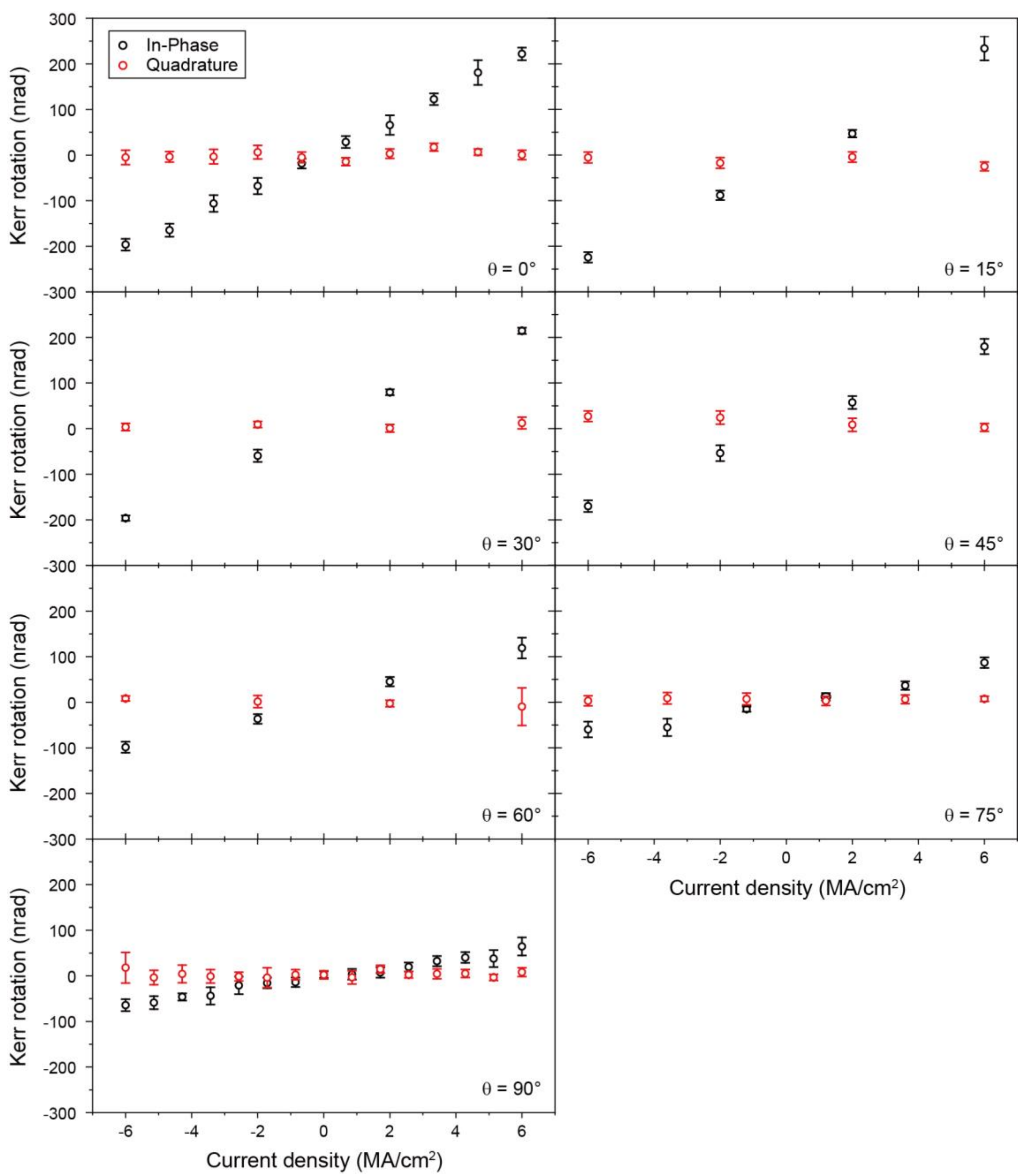

**Fig. S5. Angular dependence of current-induced MOKE signals.** Current dependence of MOKE signal measured for devices with different orientations with respect to [010] direction, θ.

## 7. Estimation of spin accumulation from MOKE measurements

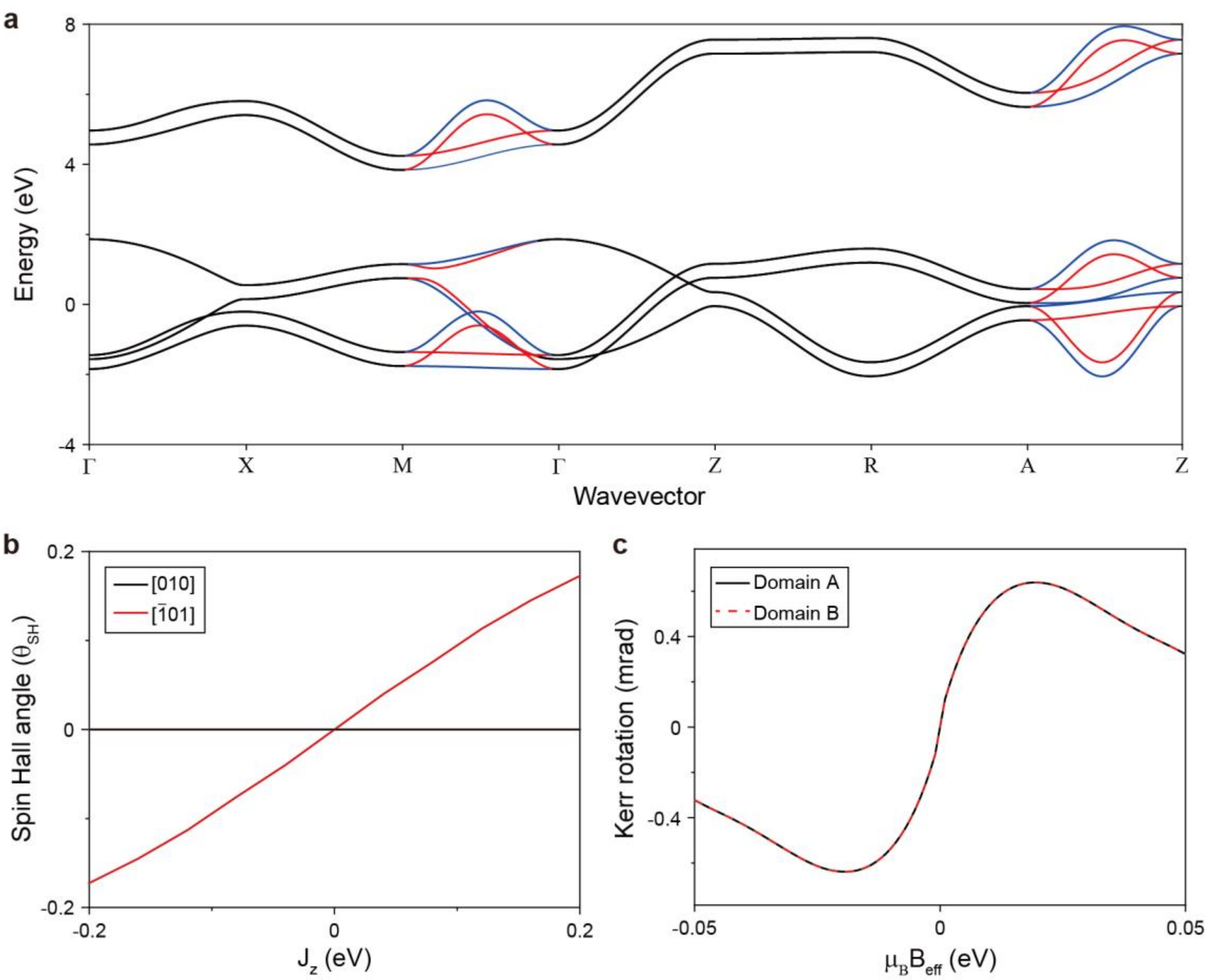

**Fig. S6. Band structure and Kerr rotation. a,** Spin-resolved band structure of $RuO_2$. Red and blue lines represent energy bands for spin-up and spin-down electrons, respectively. Black lines indicate spin-degenerate bands. **b,** Estimated spin Hall angle for currents along the [010] and $[\bar{1}01]$ directions as a function of Néel order strength. **c,** Estimated polar Kerr rotation from the (101)-oriented $RuO_2$ surface as a function of the effective magnetic field.

In this section, we detail the theoretical framework used to quantify the relevant spintronic parameters from our MOKE measurements. We first calculated the spin-resolved electronic band structure of $RuO_2$ using a 12-band tight-binding model (Fig. S6a), as described in Ref. [7]. The strength of the Néel order was set to $J_z = 0.2$ eV. In the resulting band structure, the red and blue curves denote the energy bands for spin-up and spin-down electrons, respectively, while black curves indicate regions of spin degeneracy. Notably, significant spin-split bands emerge along the M-Γ and A-Z symmetry lines, with splitting magnitudes on the order of eV.

Next, we discuss the domain-dependent spin accumulation through ASSE by computing the spin Hall angle defined as

$$\theta = \frac{2e}{\hbar}\frac{J_S}{J_C},$$

where $J_s$ and $J_c$ denote the spin and charge currents, respectively. These quantities are computed using the Boltzmann formalism as follows:

$$J_{S,ij}^{\alpha} = \frac{e\tau}{V_c\hbar}E\left[\frac{1}{N}\sum_{nk} J_{i,nk}^{\alpha} V_{j,nk}\left(-\frac{\partial f_{nk}}{\partial \epsilon}\right) d_j\right],$$

$$J_{C,ij}^{\alpha} = \frac{e^2\tau}{V_c\hbar^2}E\left[\frac{1}{N}\sum_{nk} V_{i,nk} V_{j,nk}\left(-\frac{\partial f_{nk}}{\partial \epsilon}\right) d_j\right].$$

Here, $V_c$ is the real-space unit cell volume, and $N$ is the number of cells in the Brillouin zone (80 × 80 × 80 = 512,000). Also, $\hat{V}_j = \partial_{k_j}\hat{H}$ is the derivative of Hamiltonian, $\hat{J}_i^{\alpha} = \frac{1}{2}\{\hat{V}_i, \hat{s}^{\alpha}\}$ with Pauli matrices $\hat{s}^{\alpha}$ is the spin current operator, $f_{n,k}$ is the Fermi-Dirac distribution, $\boldsymbol{E} = E\boldsymbol{d}$ is the electric field with the driving direction $\boldsymbol{d}$, and $O_{nk}$ is the expectation value $\langle nk|\hat{O}|nk\rangle$ of an operator $\hat{O}$. Figure S6b shows the calculated spin Hall angle as a function of Néel order strength varying from −0.2 to 0.2 eV. The spin Hall angle is finite solely for current injected along the [010] direction, reaching a magnitude of ~0.17 at 0.2 eV. Notably, the response exhibits a sign reversal upon the inversion of $J_z$, indicating that domains A and B host opposite surface spin accumulation.

Since the MOKE signal is directly related to the complex optical conductivity tensor, we calculated the conductivity components $\sigma_{\alpha\beta}(\omega)$ using the Kubo formula, which explicitly accounts for interband contributions:

$$\sigma_{\alpha\beta}(\omega) = \frac{e^2\hbar}{iV}\sum_{\boldsymbol{k}}\sum_{n\neq m}\frac{f_{n,\boldsymbol{k}} - f_{m,\boldsymbol{k}}}{\varepsilon_{n,\boldsymbol{k}} - \varepsilon_{m,\boldsymbol{k}}}\frac{\langle n|v_\alpha|m\rangle\langle m|v_\beta|n\rangle}{\hbar\omega + \varepsilon_{n,\boldsymbol{k}} - \varepsilon_{m,\boldsymbol{k}} + i\eta}.$$

In this expression, $e$ is the elementary charge, $\hbar$ is the reduced Planck constant, and $V$ is the crystal volume. The indices $n$ and $m$ refer to electronic bands with energies $\varepsilon_{n,\boldsymbol{k}}$ and occupations $f_{n,\boldsymbol{k}}$, while $v_\alpha$ denotes the velocity operator along the $\alpha$ direction. $\eta$ is a small broadening parameter associated with the electronic scattering rate. The calculation was performed at $\hbar\omega$ = 2.40 eV, matching the experimental laser energy. The MOKE amplitude was then determined using

$$\theta_K + i\eta_K = -\frac{2\sigma_{xy}}{(\sigma_{xx})\sqrt{1+\frac{i}{\omega\varepsilon_0}\sigma_{xx}}},$$

where $\theta_K$ and $\eta_K$ represent the Kerr rotation and ellipticity, respectively.

To translate the measured MOKE signals into an equivalent spin accumulation, we simulated the Kerr rotation as a function of an effective magnetic field (Fig. S6c). The calculated MOKE amplitude peaks at 0.6 mrad at an effective field of 0.02 eV and then gradually decreases at higher field strengths. The two antiparallel altermagnetic domains (A and B) exhibit the same magnitudes of Kerr rotation under the same effective field. Because ASSE applies opposite effective fields to domains with antiparallel Néel vectors, the resulting MOKE signals are equal in magnitude but opposite in sign. Given that the experimental spin accumulation lies within the linear-response regime, well below the field range plotted in Fig. S6c, we further evaluated the derivative of the Kerr rotation with respect to the effective magnetic field. The Kerr rotation per effective magnetic field reaches ~0.122 rad/eV near zero field. From the MOKE measurements, we estimate the effective field strength to be ~141.8 G, a magnitude that cannot be accounted for solely by the Oersted field.

**8. 4D-STEM analysis of structural heterogeneity of (101) $RuO_2$ film**

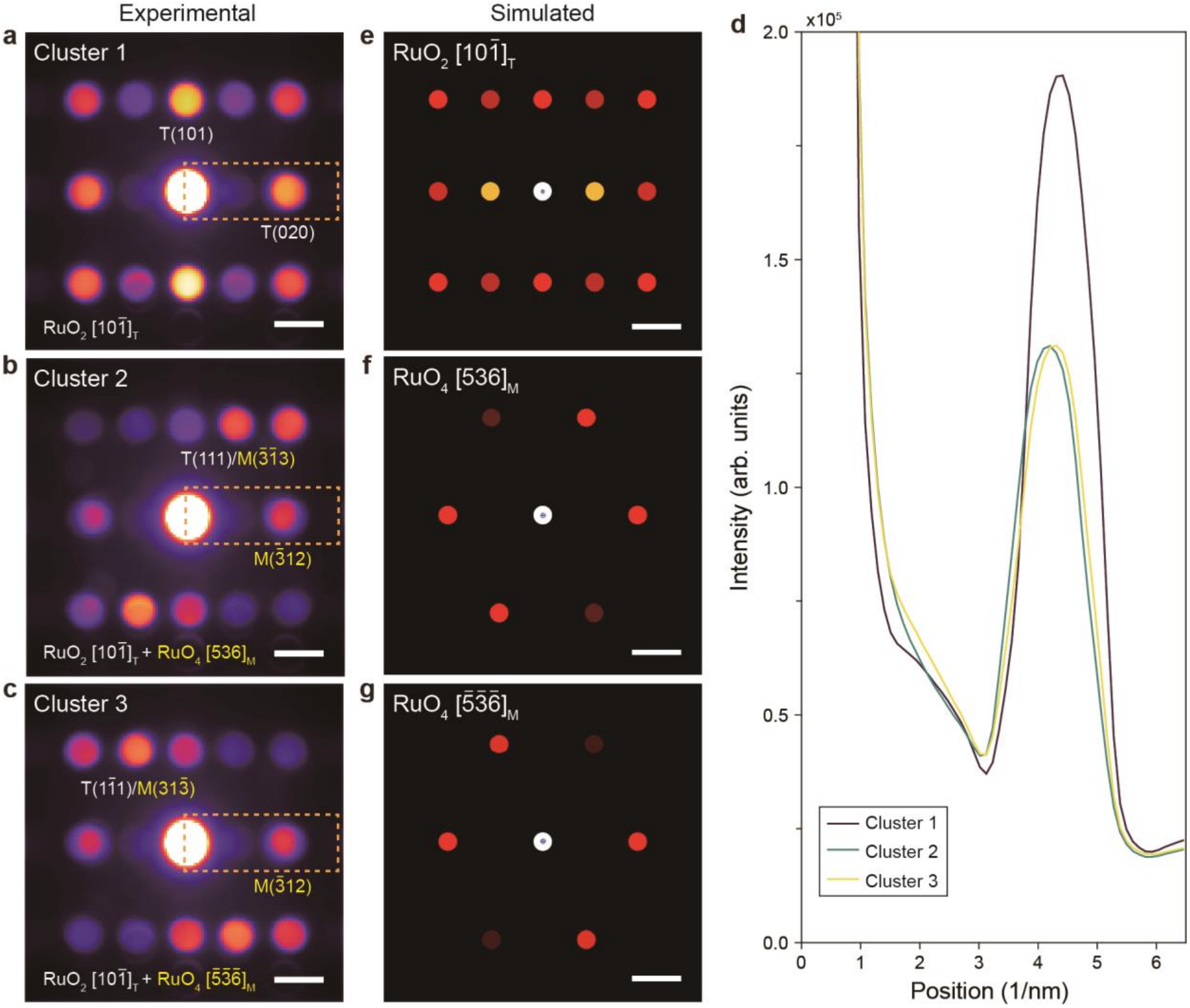

**Fig. S7. 4D-STEM analysis of structural heterogeneity. a-c,** Averaged NBED patterns acquired from Clusters 1 (**a**), 2 (**b**), and 3 (**c**). Scale bar, 2 nm$^{-1}$. **d,** Averaged line profiles of the NBED intensity for each cluster, extracted from the highlighted regions in **a-c**. **e-g,** Simulated NBED patterns corresponding to the tetragonal $RuO_2$ [10$\bar{1}$] zone axis (**e**) and the monoclinic $RuO_4$ [536] (**f**) and [$\bar{5}\bar{3}\bar{6}$] zone axis (**g**). Scale bar, 2 nm$^{-1}$.

In this section, we detail the structural classification of the three observed clusters in the $RuO_2$ thin film by comparing experimental nanobeam electron diffraction (NBED) patterns with simulated patterns for candidate ruthenium oxide phases. Cluster 1 (Fig. S7a) exhibits diffraction features consistent with the simulated diffraction pattern of (101)-oriented tetragonal $RuO_2$ viewed along the $[10\bar{1}]$ zone axis (Fig. S7e), confirming the epitaxial relationship between the (101)-oriented film and the *r*-plane sapphire substrate. In contrast, Clusters 2 and 3 exhibit measurable lateral shifts of diffraction disks along the central horizontal axis compared to Cluster 1 (Fig. S7b-c). These shifts are clearly quantified in the intensity profiles shown in Fig. S7d.

Together with the measured disk-position shifts, the distinct overall intensity distributions of the NBED patterns suggest that Clusters 2 and 3 involve local structural variations beyond simple elastic distortion of the tetragonal $RuO_2$ lattice. In particular, Clusters 2 and 3 exhibit enhanced diffraction intensity associated with the $(\bar{3}\bar{1}3)$ and $(31\bar{3})$ disks of the monoclinic $RuO_4$ phase. These features are consistent with the simulated NBED patterns of monoclinic $RuO_4$ viewed along the [536] zone axis for Cluster 2 and the $[\bar{5}\bar{3}\bar{6}]$ zone axis for Cluster 3 (Fig. S7f-g). These results indicate that, while the majority of the film maintains the pristine (101)-oriented tetragonal $RuO_2$ structure, local structural variations associated with $RuO_x$-related monoclinic-like phases are present. The coexistence of these structurally distinct regions suggests that the observed nanoscale strain may originate from lattice-mismatch accommodation at their boundaries.

**Supplementary references**